# Clause-Learning Algorithms with Many Restarts and Bounded-Width Resolution

**Albert Atserias**                                     ATSERIAS@LSI.UPC.EDU
*Universitat Politècnica de Catalunya*
*Barcelona, Spain*

**Johannes Klaus Fichte**                               FICHTE@KR.TUWIEN.AC.AT
*Vienna University of Technology*
*Vienna, Austria*

**Marc Thurley**                                        MARC.THURLEY@GOOGLEMAIL.COM
*University of California at Berkeley*
*Berkeley, USA*

## Abstract

We offer a new understanding of some aspects of practical SAT-solvers that are based on DPLL with unit-clause propagation, clause-learning, and restarts. We do so by analyzing a concrete algorithm which we claim is faithful to what practical solvers do. In particular, before making any new decision or restart, the solver repeatedly applies the unit-resolution rule until saturation, and leaves no component to the mercy of non-determinism except for some internal randomness. We prove the perhaps surprising fact that, although the solver is not explicitly designed for it, with high probability it ends up behaving as width-$k$ resolution after no more than $O(n^{2k+2})$ conflicts and restarts, where $n$ is the number of variables. In other words, width-$k$ resolution can be thought of as $O(n^{2k+2})$ restarts of the unit-resolution rule with learning.

## 1. Introduction

The discovery of a method to introduce practically feasible clause learning and non-chronological backtracking to DPLL-based solvers layed the foundation of what is sometimes called "modern" SAT-solving (Silva & Sakallah, 1996; Bayardo & Schrag, 1997). These methods set the ground for new effective implementations (Moskewicz, Madigan, Zhao, Zhang, & Malik, 2001) that spawned tremendous gains in the efficiency of SAT-solvers with many practical applications. Such great and somewhat unexpected success seemed to contradict the widely assumed intractability of SAT, and at the same time uncovered the need for a formal understanding of the capabilities and limitations underlying these methods.

Several different approaches have been suggested in the literature for developing a rigorous understanding. Among these we find the proof-complexity approach, which captures the power of SAT-solvers in terms of propositional proof systems (Beame, Kautz, & Sabharwal, 2003, 2004; Hertel, Bacchus, Pitassi, & Gelder, 2008; Pipatsrisawat & Darwiche, 2009), and the rewriting approach, which provides a useful handle to reason about the properties of the underlying algorithms and their correctness (Nieuwenhuis, Oliveras, & Tinelli, 2006). In both approaches, SAT-solvers are viewed as algorithms that search for proofs in some underlying proof system for propositional logic. With this view in mind, it was illuminating to understand that the proof system underlying modern solvers is always a subsystem of resolution (Beame et al., 2003). In particular, this means that their performance can never beat resolution lower bounds, and at the same time it provides many explicit examples where SAT-solvers require exponential time. Complementing this is the result that an idealized SAT-solver that relies on non-determinism to apply the techniques in the best possible way will be able to perform as good as general resolution (weak forms of this statement were first established in Beame et al., 2003, 2004; Hertel et al., 2008, and in the current form in Pipatsrisawat & Darwiche, 2009). As Beame et al. (2004) put it, the negative proof complexity results uncover examples of inherent intractability even under perfect choice strategies, while the positive proof complexity results give hope of finding a good choice strategy.

In this work we add a new perspective to this kind of rigorous result. We try to avoid non-deterministic choices on all components of our abstract solver and still get positive proof complexity results. Our main finding is that a concrete family of SAT-solvers that do not rely on non-determinism besides mild randomness is at least as powerful as bounded-width resolution. The precise proof-complexity result is that under the unit-propagation rule and a standard learning scheme considered by state-of-the-art solvers, the totally random decision strategy needs no more than $O(k^2 \ln(kn)n^{2k+1})$ conflicts and deterministic restarts to detect the unsatisfiability of any CNF formula on $n$ variables having a width-$k$ resolution refutation, with probability at least $1/2$. Remarkably, the analysis will provide an exact expression for this upper bound that holds for all values of $n$ and $k$ and in particular the bound we get is *not* asymptotic. Another remarkable feature is that our analysis is insensitive to whether the algorithm implements non-chronological backtracking or heuristic-based decisions provided it restarts often enough, and provided it performs totally random decisions often enough. Further details about this are given in Section 2.

By itself this result has some nice theoretical consequences, which we shall sketch briefly. First, although not explicitly designed for that purpose, SAT-solvers are able to solve instances of 2-SAT in polynomial time since every unsatisfiable 2-CNF formula has a resolution refutation of width two. More strongly, our result can be interpreted as showing that width-$k$ resolution can be simulated by $O(k^2 \ln(kn)n^{2k+1})$ rounds of unit-clause propagation. To our knowledge, such a tight connection between width-$k$ resolution and repeated application of "width-one" methods was unknown before. Another consequence is that SAT-solvers are able to solve formulas of bounded branch-width (and hence bounded treewidth) in polynomial time. We elaborate on this later in the paper. Finally, from the partial automatizability results of Ben-Sasson and Wigderson (1999), it follows that SAT-solvers are able to solve formulas having polynomial-size tree-like resolution proofs in quasipolynomial time, and formulas having polynomial-size general resolution proofs in subexponential time.

Concerning our techniques, it is perhaps surprising that the proof of our main result does not proceed by showing that the width-$k$ refutation is learned by the algorithm. For all we know the produced proof has much larger width. The only thing we show is that every width-$k$ clause in the refutation is *absorbed* by the algorithm, which means that it behaves as if it had been learned, even though it might not. In particular, if a literal and its complement are both absorbed, the algorithm correctly declares that the formula is unsatisfiable. This sort of analysis is the main technical contribution of this paper.

## 1.1 Related Work

The first attempt to compare the power of SAT-solvers with the power of resolution as a proof system was made by Beame et al. (2003, 2004). The main positive result from their work is that clause learning with a specific learning scheme and without restarts can provide exponentially shorter proofs than proper refinements of resolution such as tree, or regular, or positive resolution. Furthermore, they show that the modification of a standard solver to allow multiple assignments on the same variable would be able to simulate general resolution efficiently, assuming an ideal decision strategy. Following work showed that the requirement for multiple assignments on the same variable is a technical issue that can be avoided if the given CNF formula is pre-processed appropriately (Hertel et al., 2008). In our work we avoid these two maneuvers by introducing the concept of clause-absorption to help us analyze the standard algorithms directly.

Interestingly, for clauses that are logical consequences of the input formulas, our concept of *clause-absorption* turns out to be dual to the concept of *1-empowerment* introduced independently by Pipatsrisawat and Darwiche (2009)[1]. They used 1-empowerment to show that SAT-solvers without any conceptual modification in their operation are able to simulate general resolution efficiently, again assuming an ideal decision strategy. For comparison, our goal settles for a weaker simulation result, bounded-width resolution instead of general resolution, but does not rely on the non-determinism of ideal decision. We show that the totally random decision strategy is good enough for this purpose, provided we restart often enough. To complete this point, it is worth noting that the non-automatizability results of Alekhnovich and Razborov (2008) indicate that we cannot expect an efficient simulation of general resolution and completely avoid non-determinism at the same time.

The fact that both concepts were discovered independently adds confidence to our belief that they will play a role in subsequent studies of the power of SAT-solvers. Indeed, our techniques were recently extended to show that SAT-solvers with a totally random decision strategy are able to efficiently simulate local consistency techniques for general constraint satisfaction problems (Jeavons & Petke, 2010).

## 1.2 Organization

In Section 2 we introduce basic notation and we define the algorithm we analyze. We also discuss the dependence of our results on our choice of the learning scheme, the restart policy and the decision strategy used by the algorithm. Section 3 starts with some elementary facts about the runs of the algorithm, continues with the key definitions of *absorption* and *beneficial rounds*, and then with the analysis of the running time of the algorithm. Section 4 contains a discussion of the consequences, including the implications for formulas of bounded treewidth.

## 2. Clause Learning Algorithms

In this section we will define the algorithm and discuss our choice of its components. We start with some preliminary definitions.

---

1. Note that, originally, a weaker version of 1-empowerment was introduced by Pipatsrisawat and Darwiche (2008).

## 2.1 Preliminaries

Let $V = \{v_1, \ldots, v_n\}$ be a fixed set of propositional variables. A *literal* is a propositional variable $x$ or its negation $\bar{x}$. We use the notation $x^0$ for $\bar{x}$ and $x^1$ for $x$. Note that $x^a$ is defined in such a way that the *assignment* $x = a$ satisfies it. For $a \in \{0, 1\}$, we also use $\bar{a}$ for $1 - a$, and for a literal $\ell = x^a$ we use $\bar{\ell}$ for $x^{1-a}$. A *clause* is a set of literals, and a formula in *conjunctive normal form* (*CNF-formula*) is a set of clauses. The *width* of a clause is the number of literals in it. In the following, all formulas are over the same set of variables $V$ and every clause contains only literals on variables from $V$.

For two clauses $A = \{x, \ell_1, \ldots, \ell_r\}$ and $B = \{\bar{x}, \ell'_1, \ldots, \ell'_s\}$ we define the *resolvent of $A$ and $B$ on $x$* by $\mathrm{Res}(A, B, x) = \{\ell_1, \ldots, \ell_r, \ell'_1, \ldots, \ell'_s\}$. If the variable we *resolve on*, $x$, is implicit we simply write $\mathrm{Res}(A, B)$. A clause may contain a literal and its negation. Note that the resolvent $\mathrm{Res}(A, B, x)$ of $A$ and $B$ on $x$ is still well-defined in this case. A *resolution refutation* of a CNF formula $F$ is a sequence of clauses $C_1, \ldots, C_m$ such that $C_m = \emptyset$ and each clause $C_i$ in the sequence either belongs to $F$ or is a resolvent of previous clauses in the sequence. The *length* of a refutation is the number $m$ of clauses in the sequence. For a clause $C$, a variable $x$, and a truth value $a \in \{0, 1\}$, the *restriction* of $C$ on $x = a$ is the constant $\mathbf{1}$ if the literal $x^a$ belongs to $C$, and $C \setminus \{x^{1-a}\}$ otherwise. We write $C|_{x=a}$ for the restriction of $C$ on $x = a$.

A *partial assignment* is a sequence of assignments $(x_1 = a_1, \ldots, x_r = a_r)$ with all variables distinct. Let $\alpha$ be a partial assignment. We say that $\alpha$ *satisfies* a literal $x^a$ if it contains $x = a$. We say that $\alpha$ *falsifies* it if it contains $x = 1 - a$. If $C$ is a clause, we let $C|_\alpha$ be the result of applying the restrictions $x_1 = a_1, \ldots, x_r = a_r$ to $C$. Clearly the order does not matter. We say that $\alpha$ *satisfies* $C$ if it satisfies at least one of its literals; i.e., if $C|_\alpha = \mathbf{1}$. We say that $\alpha$ *falsifies* $C$ if it falsifies all its literals; i.e., if $C|_\alpha = \emptyset$. If $D$ is a set of clauses, we let $D|_\alpha$ denote the result of applying the restriction $\alpha$ to each clause in $D$, and removing the resulting $\mathbf{1}$'s. We call $D|_\alpha$ the *residual* set of clauses.

## 2.2 Definition of the Algorithm

A *state* is a sequence of assignments $(x_1 = a_1, \ldots, x_r = a_r)$ in which all variables are distinct and some assignments are marked as *decisions*. We use the notation $x_i \overset{\mathrm{d}}{=} a_i$ to denote that the assignment $x_i = a_i$ is a *decision assignment*. In this case $x_i$ is called a *decision variable*. The rest of assignments are called *implied assignments*. We use $S$ and $T$ to denote states. The empty state is the one without any assignments. Define the *decision level* of an assignment $x_i = a_i$ as the number of decision assignments in $(x_1 = a_1, \ldots, x_i = a_i)$. When convenient, we identify a state with the underlying partial assignment where all decision marks are ignored.

### 2.2.1 Operation

The algorithm maintains a current state $S$ and a current set of clauses $D$. There are four modes of operation DEFAULT, CONFLICT, UNIT, and DECISION. The algorithm starts in DEFAULT mode with the empty state as the current state and the given CNF formula as the current set of clauses:

- DEFAULT. If $S$ sets all variables in $D$ and satisfies all clauses in $D$, stop and output SAT together with the current state $S$. Otherwise, if $D|_S$ contains the empty clause, move to CONFLICT mode. Otherwise, if $D|_S$ contains a unit clause, move to UNIT mode. Finally, if control reaches this point, move to DECISION mode.

- CONFLICT. Apply the *learning scheme* to add a new clause $C$ to $D$. If $C$ is the empty clause, stop and output UNSAT. Otherwise, apply the *restart policy* to decide whether to continue further or to restart in DEFAULT mode with the current $D$ and $S$ initialized to the empty state. In case we continue further, repeatedly remove assignments from the tail of $S$ as long as $C|_S = \emptyset$, and then go to UNIT mode.

- UNIT. For any unit clause $\{x^a\}$ in $D|_S$, add $x = a$ to $S$ and go back to DEFAULT mode.

- DECISION. Apply the *decision strategy* to determine a decision $x \overset{\mathrm{d}}{=} a$ to be added to $S$ and go back to DEFAULT mode.

To guarantee correctness and termination, the learning scheme will always add a clause $C$ that is a logical consequence of $D$, for which $C|_S = \emptyset$ holds at the time it is added, and that contains at most one variable of maximum decision level. It is not hard to see that these properties prevent such a clause from being learned twice, and since the number of clauses on the variables of $D$ is finite, this implies termination. Clauses with these characteristics always exist as they include the *asserting clauses* (Zhang, Madigan, Moskewicz, & Malik, 2001) that will be discussed in Section 2.3.3.

The well-known DPLL-procedure is a precursor of this algorithm where, in CONFLICT mode, the learning scheme never adds any new clause, the restart policy does not dictate any restart at all, and assignments are removed from the tail of $S$ up to the latest decision assignment, say $x \overset{\mathrm{d}}{=} a$, which is replaced by $x \overset{\mathrm{d}}{=} 1 - a$. We say that the DPLL-procedure *backtracks* on the latest decision. In contrast, modern SAT-solvers implement learning schemes and backtrack on a literal, as determined by the learned clause, which is not necessarily the latest decision. This is called *non-chronological backtracking*. Besides learning schemes and non-chronological backtracking, modern SAT-solvers also implement restart policies and appropriate decision strategies. We discuss our choice of these components of the algorithm in Section 2.3.

### 2.2.2 Runs of the Algorithm

Consider a run of the algorithm started in DEFAULT mode with the empty state and initial set of clauses $D$, until either a clause is falsified or all variables are set. Such a run is called a *complete round started with $D$* and we represent it by the sequence of states $S_0, \ldots, S_m$ that the algorithm goes through, where $S_0$ is the empty state and $S_m$ is the state where either all variables are set, or the falsified clause is found. More generally, a *round* is an initial segment $S_0, \ldots, S_r$ of a complete round up to a state where either $D|_{S_r}$ contains the empty clause or $D|_{S_r}$ does not contain any unit clause. If $D|_{S_r}$ contains the empty clause we say that the round is *conclusive*. If a round is not conclusive we call it *inconclusive*. The term inconclusive means to reflect the fact that no clause can be learned from such a round. In particular, a (complete) round that ends in a satisfying assignment is inconclusive[2].

For a round $S_0, \ldots, S_r$, note that for $i \in \{1, \ldots, r\}$, the state $S_i$ extends $S_{i-1}$ by exactly one assignment of the form $x_i = a_i$ or $x_i \stackrel{\mathrm{d}}{=} a_i$ depending on whether UNIT or DECISION is executed at that iteration; no other mode assigns variables. When this does not lead to confusion, we identify a round with its last state interpreted as a partial assignment. In particular, we say that the round *satisfies* a clause $C$ if $C|_{S_r} = \mathbf{1}$, and that it *falsifies* it if $C|_{S_r} = \emptyset$.

## 2.3 Restart Policy, Learning Scheme, and Decision Strategy

In the following we will discuss our choice of the learning scheme, the restart policy and the decision strategy used by the algorithm. Our discussion will particularly focus on the dependence of our results on this choice.

### 2.3.1 RESTART POLICY

The *restart policy* determines whether to restart the search after a clause is learned. The only important characteristic that we need from the restart policy is that it should dictate restarts often enough. In particular, our analysis will work equally well for the most aggressive of the restart policies, the one that dictates a restart after every conflict, as for a less aggressive strategy that allows any bounded number of conflicts between restarts. The fact that our analysis is insensitive to this will follow from a monotonicity property of the performance of the algorithm that we will prove in Lemma 5. More precisely, it will follow from the monotonicity lemma that if we decide to use a policy that allows $c > 1$ conflicts before a restart, then the upper bound on the number of required restarts can only decrease (or stay the same). Only the upper bound on the number of conflicts would appear multiplied by a factor of $c$, even though the truth might be that even those decrease as well. For simplicity of exposition, for the rest of the paper we assume that the restart policy dictates a restart after every conflict.

### 2.3.2 DECISION STRATEGY

The *decision strategy* determines which variable is assigned next, and to what value. Again, the only important characteristic that we need from the decision strategy is that it should allow a round of totally random decisions often enough. Here, a totally random decision is defined as follows: if the current state of the algorithm is $S$, we choose a variable $x$ uniformly at random among the variables from $V$ that do not appear in $S$, and a value $a$ in $\{0, 1\}$ also uniformly at random and independently of the choice of $x$. Thus, our analysis actually applies to any decision strategy that allows any bounded number of rounds with heuristic-based decisions between totally random ones. More precisely, if we allow say $c > 1$ rounds of non-random decisions between random ones, then the number of required restarts and conflicts would appear multiplied by a factor of $c$. Again this will follow from the monotonicity lemma referred to above. That said, for simplicity of exposition we assume in the following that every decision is totally random.

---

2. Let us note that the definitions of round, conclusive round and inconclusive round differ slightly from those given in the conference version of this paper (Atserias, Fichte, & Thurley, 2009). The current definitions make the concepts more robust.

### 2.3.3 Learning Scheme

The *learning scheme* determines which clause will be added to the set of clauses when a conflict occurs. Let $S_0, \ldots, S_r$ be a conclusive round started with the set of clauses $D$ that ends up falsifying some clause of $D$. Let $x_i = a_i$ or $x_i \overset{\mathrm{d}}{=} a_i$ be the $i$-th assignment of the round. We annotate each $S_i$ by a clause $A_i$ by reverse induction on $i \in \{1, \ldots, r\}$:

1. Let $A_{r+1}$ be any clause in $D$ that is falsified by $S_r$.

2. For $i \leq r$ for which $x_i \overset{\mathrm{d}}{=} a_i$ is a decision, let $A_i = A_{i+1}$.

3. For $i \leq r$ for which $x_i = a_i$ is implied, let $B_i$ be any clause in $D$ such that $B_i|_{S_{i-1}}$ is the unit clause $\{x_i^{a_i}\}$, and let $A_i = \mathrm{Res}(A_{i+1}, B_i, x_i)$ if these clauses are resolvable on $x_i$, and let $A_i = A_{i+1}$ otherwise.

It is quite clear from the construction that each $A_i$ has a resolution proof from the clauses in $D$. In fact, the resolution proof is linear and even trivial in the sense of Beame et al. (2004). We call each clause $A_i$ a *conflict clause*. If $d$ denotes the maximum decision level of the assignments in $S_r$, a conflict clause is called an *asserting clause* if it contains exactly one variable of decision level $d$. Asserting clauses, originally defined by Zhang et al. (2001), capture the properties of conflict clauses learned by virtually any modern SAT-solver. For brevity, we describe only two concrete learning schemes in detail. For other schemes see the work of Zhang et al. (2001).

The Decision learning scheme adds clause $A_1$ to the current set of clauses after each conflict. It is not hard to check that $A_1$ is an asserting clause. Furthermore, every literal in $A_1$ is the negation of some decision literal in $S_r$; this will be important later on. The 1UIP learning scheme, which stands for *1st Unique Implication Point*, is the one that adds a clause $A_i$ such that $i \leq r$ is maximal subject to the condition that $A_i$ is an asserting clause.

In the following we will assume, tacitly, that the algorithm employs *some* asserting learning scheme, that is, one whose learned clauses are always asserting, except for the empty clause.

### 2.3.4 Clause Bookkeeping

It should be mentioned that our analysis relies crucially on the assumption that the learned clauses are never removed from the current set of clauses. However, practical SAT-solvers periodically delete some of the learned clauses to save memory and to avoid the overhead they introduce. Thus an interesting question is whether our results can be made to work without the assumption. In this respect, the strong proof-complexity results of Nordström (2009) showing that not every small-width resolution refutation can be made to work in small clause-space seems to indicate that an assumption similar to ours is indeed needed.

Another remark worth making at this point concerns the width of the learned clauses. Since our goal is to show that the algorithm can simulate small-width resolution, it seems natural to ask whether we can restrict the learning scheme to learn clauses of small width only. As mentioned in the introduction, our analysis does not seem to allow it. Moreover, recent results by Ben-Sasson and Johannsen (2010) show that, in general, learning short clauses only is a provably weaker scheme than learning arbitrarily long clauses. Thus, while the examples of Ben-Sasson and Johannsen (2010) do not have small-width resolution refutations and therefore do not show that keeping long clauses is actually required in this case, it is conceivable that it might.

## 3. Analysis of the Algorithm

In this section we will analyze the running time of the algorithm. Before we can do this, however, we will have to introduce our key technical concepts of absorption and beneficial rounds, and study some of their most important properties.

### 3.1 Runs of the Algorithm

Let $R$ and $R'$ be rounds, and let $C$ be a clause. We say that $R'$ *subsumes* $R$ if, up to decision marks, every assignment in $R$ appears also in $R'$. We say that $R$ and $R'$ *agree* on $C$ if the restrictions of $R$ and $R'$ to variables in $C$ are equal: every variable in $C$ is either unassigned in both, or assigned to the same value in both. We say that $R$ *branches* in $C$ if all decision variables of $R$ are variables in $C$. Note that the properties *agree on $C$* and *branches in $C$* depend only on the set of variables of $C$. We define them for clauses to simplify notation later on.

We prove two rather technical lemmas. The goal is to show that inconclusive rounds are robust with respect to the order in which assignments are made. For example, the first lemma shows that any inconclusive round subsumes any other round that agrees with it on its decisions. In fact we will need a slightly stronger claim that involves rounds from two different sets of clauses.

**Lemma 1.** *Let $D$ and $D'$ be sets of clauses with $D \subseteq D'$, let $C$ be a clause, and let $R'$ be an inconclusive round started with $D'$. Then, for every round $R$ started with $D$ that branches in $C$ and agrees with $R'$ on $C$, it holds that $R'$ subsumes $R$.*

*Proof.* Let $R = (S_0, \ldots, S_r)$. By induction on $i$, we prove that for every $i \in \{0, \ldots, r\}$, every assignment in $S_i$ is also made in $R'$. For $i = 0$ there is nothing to prove since $S_0 = \emptyset$. Let $i > 0$ and assume that every assignment in $S_{i-1}$ is also made in $R'$. Let $x = a$ or $x \stackrel{\mathrm{d}}{=} a$ be the last assignment in $S_i$. Since $R$ and $R'$ agree on $C$ and $R$ branches in $C$, every decision assignment made in $R$ is also made in $R'$. This takes care of the case $x \stackrel{\mathrm{d}}{=} a$. Suppose then that the last assignment $x = a$ in $S_i$ is implied. This means that there exists a clause $A$ in $D$ such that $A|_{S_{i-1}} = \{x^a\}$. Since $D \subseteq D'$ and every assignment made in $S_{i-1}$ is also made in $R'$, necessarily $x = a$ appears in $R'$ because $R'$ is inconclusive and cannot leave unit clauses unset. □

The next lemma shows that the universal quantifier in the conclusion of the previous lemma is not void. In addition, the round can be chosen inconclusive.

**Lemma 2.** *Let $D$ and $D'$ be sets of clauses with $D \subseteq D'$, let $C$ be a clause, and let $R'$ be an inconclusive round started with $D'$. Then, there exists an inconclusive round $R$ started with $D$ that branches in $C$ and agrees with $R'$ on $C$, and such that $R'$ subsumes $R$.*

*Proof.* Let $R' = (T_0, \ldots, T_t)$. Define $I \subseteq \{0, \ldots, t\}$ as the set of indices $i$ such that the $i$-th assignment of $R'$ assigns some variable in $C$. For $i \in I$, let $x_i = a_i$ or $x_i \stackrel{\mathrm{d}}{=} a_i$ be the $i$-th assignment in $R'$.

We will construct a round $R = (S_0, \ldots, S_s)$ started with $D$ inductively. Associated with each $S_j$ is the set $I_j \subseteq I$ of indices $i$ such that $x_i$ is left unassigned in $S_j$. Recall that $S_0$ is the empty state by definition. Hence $I_0 = I$. We define the following process:

1. If $S_j$ falsifies some clause in $D$ or it sets all variables in $V$ then set $s = j$ and stop.

2. Otherwise, if there is a unit clause $\{x^a\}$ in $D|_{S_j}$ then let $S_{j+1}$ be $S_j$ plus $x = a$.

3. Otherwise, if $I_j$ is non-empty, let $i$ be the minimum element of $I_j$, and let $S_{j+1}$ be obtained by adding the decision $x_i \stackrel{\mathrm{d}}{=} a_i$ to $S_j$.

If none of the above cases applies set $s = j$ and stop the process.

By construction $R$ is a valid round started with $D$. Let us see that $R'$ subsumes $R$: let $A$ be the set of literals made true by decisions in $R$. By construction, $R$ and $R'$ agree on $A$ and hence $R'$ subsumes $R$ by Lemma 1. Furthermore, $R$ is inconclusive: By $D \subseteq D'$ and $R'$ being inconclusive, $D|_{R'}$ does not contain the empty clause, and as $R'$ subsumes $R$, also $D|_R$ does not contain the empty clause. Further, as every variable in $C$ belongs to $V$ and $R$ is inconclusive, the process stops with $I_s = \emptyset$. Together with the fact that $R'$ subsumes $R$, this shows that $R$ and $R'$ agree on $C$. Note finally that $R$ branches in $C$ by construction. □

## 3.2 Absorption

One key feature of the definition of a round is that if it is inconclusive, then the residual set of clauses does not contain unit clauses and, in particular, it is *closed* under unit propagation. This means that for an inconclusive round $R$ started with $D$, if $A$ is a clause in $D$ and $R$ falsifies all its literals but one, then $R$ must satisfy the remaining literal, and hence $A$ as well. Besides those in $D$, other clauses may have this property, which is important enough to deserve a definition:

**Definition 3** (Absorption). *Let $D$ be a set of clauses, let $A$ be a non-empty clause and let $x^a$ be a literal in $A$. We say that $D$ absorbs $A$ at $x^a$ if every inconclusive round started with $D$ that falsifies $A \setminus \{x^a\}$ assigns $x$ to $a$. We say that $D$ absorbs $A$ if $D$ absorbs $A$ at every literal in $A$.*

Naturally, when $D$ absorbs $A$ at $x^a$ we also say that $A$ *is absorbed by $D$ at $x^a$*.

Intuitively, one way to think of absorbed clauses is as being learned implicitly. The rest of this section is devoted to make this intuition precise. For now, let us note that if there are no inconclusive rounds started with $D$, then every clause is absorbed. This agrees with the given intuition since the absence of inconclusive rounds means that unit-clause propagation applied on $D$ produces the empty clause. In this section we also show that the notion of *clause-absorption* is tightly connected to the concept of *1-empowerment* independently introduced by Pipatsrisawat and Darwiche (2009).

### 3.2.1 PROPERTIES OF ABSORPTION

Before we continue, let us discuss some key properties of absorption. We argued already that every clause in $D$ is absorbed by $D$. We give an example showing that $D$ may absorb other clauses. Let $D$ be the set consisting of the three clauses

$$a \vee \bar{b} \qquad b \vee c \qquad \bar{a} \vee \bar{b} \vee d \vee e.$$

In this example, the clause $a \vee c$ does not belong to $D$ but is absorbed by $D$ since every inconclusive round that sets $a = 0$ must set $c = 1$ by unit-propagation, and every inconclusive round that sets $c = 0$ must set $a = 1$ also by unit-propagation. While $D$ may absorb other clauses as we just saw, we note that every non-empty clause absorbed by $D$ is a logical consequence of $D$. We write $D \models C$, if every satisfying assignment of $D$ satisfies $C$.

**Lemma 4.** *Let $D$ be a set of clauses and let $C$ be a non-empty clause. If $D$ absorbs $C$, then $D \models C$.*

*Proof.* Let $S$ be a full assignment that satisfies all clauses in $D$. We want to show that $S$ satisfies $C$ as well. Let $R = (S_0, \ldots, S_r)$ be a complete round of the algorithm started with $D$ that sets all its decision variables as they are set in $S$. By induction on $i \in \{0, \ldots, r\}$, we will show that $S_i \subseteq S$ and it will follow that $R$ is not stopped by a conflict and therefore $S_r = S$. In particular $R$ is inconclusive, and if it falsifies all literals of $C$ but one, it must satisfy the remaining one because $C$ is absorbed. Since $R$ sets all variables in $C$ and $S_r = S$, this means that $S$ satisfies $C$.

It remains to show that $S_i \subseteq S$ for every $i$. For $i = 0$ there is nothing to show since $S_0 = \emptyset$. Fix $i > 0$ and assume that $S_{i-1} \subseteq S$. Let $x = a$ or $x \overset{\mathrm{d}}{=} a$ be the last assignment in $S_i$. The case $x \overset{\mathrm{d}}{=} a$ is taken care by the assumption that all decision variables of $R$ are set as in $S$. Suppose then that the last assignment $x = a$ is implied. This means that there exists a clause $A$ in $D$ such that $A|_{S_{i-1}} = \{x^a\}$. Since $S$ satisfies $D$ and $S_{i-1} \subseteq S$, necessarily $x$ is set to $a$ in $S$. $\qquad\square$

Next, let us see that the converse of the above lemma does not hold; namely, we see that not every implied clause is absorbed. In the previous example, for instance, note that $\bar{b} \vee d \vee e$ is a consequence of $D$ (resolve the first and the third clause on $a$) but is not absorbed by $D$ (consider the inconclusive round $d \overset{\mathrm{d}}{=} 0, e \overset{\mathrm{d}}{=} 0$).

One interesting property that is illustrated by this example is that if $C$ is the resolvent of two absorbed clauses $A$ and $B$, and $C$ is not absorbed at some literal $\ell$, then $\ell$ appears in both $A$ and $B$. In the example above, $D$ does not absorb $\bar{b} \vee d \vee e$ at $\bar{b}$, and $\bar{b}$ appears in the clauses $a \vee \bar{b}$ and $\bar{a} \vee \bar{b} \vee d \vee e$ from $D$, whose resolvent is precisely $\bar{b} \vee d \vee e$. We will prove this general fact in the next section where the objects of study will be non-absorbed resolvents of absorbed clauses.

Next we show three key monotonicity properties of clause-absorption, where the first is the one that motivated its definition.

**Lemma 5.** *Let $D$ and $E$ be sets of clauses and let $A$ and $B$ be non-empty clauses. The following hold:*

    *1. if $A$ belongs to $D$, then $D$ absorbs $A$,*

*2. if $A \subseteq B$ and $D$ absorbs $A$, then $D$ absorbs $B$,*

*3. if $D \subseteq E$ and $D$ absorbs $A$, then $E$ absorbs $A$.*

*Proof.* To prove 1. assume for contradiction that there is a literal $\ell$ in $A$ and an inconclusive round $S_0, \ldots, S_r$ started with $D$ which falsifies $A \setminus \{\ell\}$ but does not satisfy $A$. As the round is inconclusive, we cannot have $A|_{S_r} = \emptyset$, which means then that $A|_{S_r} = \{\ell\}$, in contradiction to the definition of round.

For the proof of 2. let $\ell$ be a literal of $B$ and define $B' = B \setminus \{\ell\}$. We consider two different cases. If $\ell \notin A$ then $A \subseteq B'$ and, as $A$ is absorbed by $D$, there is no inconclusive round which falsifies $B'$. Thus $B$ is absorbed in this case. If $\ell \in A$, let $A' = A \setminus \{\ell\}$ and let $S_0, \ldots, S_r$ be an inconclusive round started with $D$ which falsifies $B'$. Then it falsifies $A'$ and satisfies $A$ by absorption. Thus it satisfies $B$, and $B$ is absorbed in this case as well.

It remains to prove 3. Let $\ell$ be some literal in $A$ and $A' = A \setminus \{\ell\}$. Let $R'$ be an inconclusive round started with $E$ which falsifies $A'$. By Lemma 2, there is an inconclusive round $R$ started with $D$ which falsifies $A'$ and which is subsumed by $R'$. As $A$ is absorbed by $D$, we see that $R$ (and hence $R'$) satisfies $A$. □

### 3.2.2 ABSORPTION AND EMPOWERMENT

Our next goal is to show that absorption and empowerment are dual notions. For assignments $\alpha, \beta$ we write $\alpha \subseteq \beta$ if every assignment in $\alpha$ is also in $\beta$. Let us reproduce the definition of 1-empowerment in the work of Pipatsrisawat and Darwiche (2009), slightly adapted to better suit our notation and terminology.

**Definition 6** (1-Empowerment in Pipatsrisawat & Darwiche, 2009)**.** *Let $D$ be a set of clauses, let $C$ be a non-empty clause and let $x^a$ be a literal in $C$. Let $\alpha$ be the assignment that sets $y = 1 - b$ for every literal $y^b$ in $C \setminus \{x^a\}$. We say that $C$ is 1-empowering via $x^a$ with respect to $D$, if the following three conditions are met:*

   *1. $C$ is a logical consequence of $D$; i.e. $D \models C$,*

   *2. repeated applications of unit-clause propagation on $D|_\alpha$ do not yield the empty clause,*

   *3. repeated applications of unit-clause propagation on $D|_\alpha$ do not assign $x$ to $a$.*

*We also say that $x^a$ is an empowering literal of $C$. We say that $C$ is 1-empowering if it is 1-empowering via some literal in $C$.*

A preliminary version of this definition was given by Pipatsrisawat and Darwiche (2008) where the second of the three conditions was not required.

By the definition of absorption, we see that if some non-empty clause $A$ is *not* absorbed by a set of clauses $D$, then there is an inconclusive round $R$ started with $D$ and a literal $x^a$ in $A$ such that $R$ falsifies $A \setminus \{x^a\}$ but does not satisfy $\{x^a\}$. When $A$ is a logical consequence of $D$, this witnesses precisely the fact that $A$ is 1-empowering via $x^a$. We show that the converse is also true:

**Lemma 7.** *Let $D$ be a set of clauses, let $C$ be a non-empty clause such that $D \models C$, and let $x^a$ be a literal in $C$. Then, $C$ is 1-empowering via $x^a$ with respect to $D$ if and only if $D$ does not absorb $C$ at $x^a$.*

*Proof.* Let $C' = C \setminus \{x^a\}$. Assume first that $D$ does not absorb $C$ at $x^a$. Let $R = (S_0, \ldots, S_r)$ be an inconclusive round started with $D$ witnessing this fact, i.e. $S_r$ falsifies $C'$ and does not assign $x = a$. In particular $\alpha \subseteq S_r$. Furthermore, for every unit clause $\{y^b\}$ in $D|_\alpha$ we have $y = b$ in $S_r$, as $R$ is an inconclusive round. By a straightforward induction, we see that every $\beta$ obtained from $\alpha$ by repeated applications of unit-clause propagation from $D|_\alpha$ also satisfies $\beta \subseteq S_r$. This directly implies conditions 2. and 3. in the definition of 1-empowerment. Condition 1. is met by assumption.

For the converse, assume that $C$ is 1-empowering via $x^a$ with respect to $D$. We have to show that there is an inconclusive round started with $D$ that falsifies $C'$ but does not assign $x = a$. Let $R = (S_0, \ldots, S_r)$ be a round started with $D$ in which every decision assignment is chosen to falsify a literal in $C'$, and that, among all rounds with this property, assigns as many literals from $C'$ as possible. Clearly such a maximal round exists since the one that does not make any decision meets the property.

We shall show that $R$ is the round we seek. For each $i \in \{0, \ldots, r\}$, let $\alpha_i \subseteq \alpha$ be the maximal assignment such that $\alpha_i \subseteq S_i$, let $\beta_i$ be obtained from $\alpha_i$ by repeated applications of unit-clause propagation from $D|_{\alpha_i}$, and let $\gamma_i$ be the subset of assignments in $\beta_i$ that are also in $S_i$. In particular $\gamma_i \subseteq S_i$. We shall prove, by induction on $i$, that $S_i \subseteq \gamma_i$ and hence $S_i = \gamma_i$.

The base case $i = 0$ is trivial since $S_0 = \emptyset$. Assume now that $i > 0$ and $S_{i-1} \subseteq \gamma_{i-1}$. If the $i$-th assignment of $S_i$ is a decision assignment, then by construction it falsifies a literal in $C'$ and hence belongs to $\alpha$. But then it also belongs to $\alpha_i$, $\beta_i$ and $\gamma_i$ as required. If the $i$-th assignment of $S_i$ is implied we distinguish two cases: whether it also belongs to $\alpha$ or not. If the implied assignment is also in $\alpha$, then it is in $\alpha_i$, $\beta_i$ and $\gamma_i$ as required. If the implied assignment is not in $\alpha$, then $\alpha_i = \alpha_{i-1}$ and hence $\beta_i = \beta_{i-1}$. But then, since $S_{i-1} \subseteq \gamma_{i-1}$ by induction hypothesis and $\gamma_{i-1} \subseteq \beta_{i-1}$, the unit clause responsible for the definition of $S_i$ appears in the process of forming $\beta_{i-1}$ and hence in the process of forming $\beta_i$. Therefore the assignment will also be in $\gamma_i$.

This completes the induction and shows, in particular, that $S_r = \gamma_r$. By point 2. in the definition of 1-empowerment, $R$ is inconclusive. Furthermore, by point 3. in the definition of 1-empowerment, $S_r$ does not assign $x = a$. It remains to show that $S_r$ falsifies $C'$. First note that, by the maximality of $R$ and the fact that $R$ is inconclusive, every literal in $C'$ is assigned by $R$. Moreover, since the decision assignments of $R$ are chosen to falsify the literals in $C'$, it suffices to show that the implied assignments of $R$ do not satisfy any literal in $C'$. Thus, suppose for contradiction that $y = b$ is an implied assignment in $R$ and that $y^b$ is a literal in $C'$. Let $i \in \{0, \ldots, r\}$ be such that $\{y^b\}$ is a unit-clause in $D|_{S_i}$. Since $S_i \subseteq S_r \subseteq \gamma_r$ and $y$ is assigned to $1 - b$ in $\alpha$, the unit-clause $\{y^b\}$ in $D|_{S_i}$ appears as the empty clause in the closure under unit-clause propagation of $D|_\alpha$; this contradicts point 2. in the definition of 1-empowerment and completes the proof. □

Let us note at this point that if condition 1. in the definition of 1-empowerment is dropped, then the hypothesis that $D \models C$ can also be dropped from Lemma 7. This would make 1-empowerment and absorption literally dual of each other.

### 3.3 Beneficial Rounds

We shall now study the key situation that explains how the algorithm can possibly simulate resolution proofs. Consider the resolvent $C = \text{Res}(A, B)$ of two absorbed clauses $A$ and $B$ which itself, however, is not absorbed. Our goal is to study what $A$, $B$ and $C$ look like in such a case. We start by showing that if $C$ is not absorbed at a literal $\ell \in C$, then $\ell$ appears in both $A$ and $B$. This property held the key for discovering the concept of clause-absorption and its relevance to the simulation of resolution proofs. A similar connection to clause learning was observed by Pipatsrisawat and Darwiche (2008), where it is also pointed out that the condition that some literal from $C$ appears in both $A$ and $B$ is known as *merge resolution* (Andrews, 1968).

**Lemma 8.** *Let $D$ be a set of clauses, let $A$ and $B$ be two resolvable clauses that are absorbed by $D$, and let $C = \text{Res}(A, B)$. If $\ell$ is a literal in $C$ and $D$ does not absorb $C$ at $\ell$, then $\ell$ appears in both $A$ and $B$.*

*Proof.* Let $y$ be such that $C = \text{Res}(A, B, y)$, and let $A' = A \setminus \{y\}$ and $B' = B \setminus \{\bar{y}\}$. Let $\ell = x^a$ be a literal in $C$ and assume $D$ does not absorb $C$ at $\ell$. Then there exists an inconclusive round $R$ that falsifies $C \setminus \{x^a\}$ but does not set $x$ to $a$. Since $\ell$ belongs to $C$ and $C = A' \cup B'$ we have that $\ell$ belongs to $A$ or to $B$, or to both. If it belongs to both, we are done. Otherwise, assume without loss of generality that it belongs to $A$ but not to $B$. In this case $R$ falsifies $B \setminus \{\bar{y}\}$, and since $B$ is absorbed, $y$ is set to 0 in $R$. But then $R$ falsifies $A \setminus \{x^a\}$, and since $A$ is absorbed, $x$ is set to $a$ in $R$. This contradicts the choice of $R$ where $x$ was not set to $a$. □

We continue by showing that in the situation of interest, there always exist a *beneficial* round of the algorithm which predicts eventual absorption.

**Definition 9** (Beneficial Round). *Let $D$ be a set of clauses, let $A$ be a non-empty clause, let $x^a$ be a literal of $A$, and let $R$ be an inconclusive round started with $D$. We say that $R$ is beneficial for $A$ at $x^a$ if it falsifies $A \setminus \{x^a\}$, branches in $A \setminus \{x^a\}$, leaves $x$ unassigned, and yields a conclusive round if extended by the decision $x \overset{d}{=} \bar{a}$ . The conclusive round obtained by extending $R$ by $x \overset{d}{=} \bar{a}$ is also called* beneficial for $A$ at $x^a$. *We say that $R$ is* beneficial for $A$ if it is beneficial for $A$ at some literal in $A$.*

In other words, a round started with $D$ that is beneficial for $A$ at $x^a$ is a witness that $D$ does not absorb $A$ at $x^a$, which is minimal with this property, and yet yields a conflict when $x$ is set to the wrong value. Thus, informally, a beneficial round is a witness that $D$ *almost absorbs* $A$ at $x^a$.

**Lemma 10.** *Let $D$ be a set of clauses, let $A$ and $B$ be two resolvable clauses that are absorbed by $D$, and let $C = \text{Res}(A, B)$. If $C$ is non-empty and not absorbed by $D$, then there is a round started with $D$ that is beneficial for $C$.*

*Proof.* We identify a literal $x^a$ in $C$ for which we are able to build a beneficial round for $C$ at $x^a$.

Let $y$ be such that $C = \text{Res}(A, B, y)$, and let $A' = A \setminus \{y\}$ and $B' = B \setminus \{\bar{y}\}$. As $C$ is non-empty and not absorbed by $D$, there is a literal $x^a$ in $C$ and an inconclusive round $R'$ started with $D$ which falsifies $C' = C \setminus \{x^a\}$ but does not set $x$ to $a$. Also $x$ is not assigned $\bar{a}$ in $R'$ since otherwise it would falsify $C$, and as $C = A' \cup B'$ and $D$ absorbs both $A$ and $B$, both $y$ and $\bar{y}$ would be satisfied by $R'$. This shows that $x$ is unassigned in $R'$.

Let $R$ be the inconclusive round started with $D$ which is obtained by applying Lemma 2 to $C'$ and the given inconclusive round $R'$. We claim that $R$ is beneficial for $C$ at $x^a$: The round $R$ falsifies $C'$, as it agrees with $R'$ on $C'$. Also $R$ branches in $C'$ and, as $R'$ subsumes $R$, leaves $x$ unassigned. Finally, note that $R$ and $R'$ also agree on $A \setminus \{y\}$ and $B \setminus \{\bar{y}\}$. Hence extending the round $R$ by a decision $x \overset{\mathrm{d}}{=} \bar{a}$ yields a conclusive round; otherwise both $y$ and $\bar{y}$ would be satisfied since both $A$ and $B$ are absorbed by $D$. $\qquad\square$

## 3.4 Main Technical Lemma

We will now start analyzing the number of complete rounds it takes until the resolvent of two absorbed clauses is absorbed as a function of its width. However, as this is not trivial we first have to determine the number of complete rounds it takes until a sufficient prerequisite of absorption occurs: a beneficial round.

**Lemma 11.** *Let $D$ be a set of clauses, and let $A$ and $B$ be two resolvable clauses that are absorbed by $D$ and that have a non-empty resolvent $C = \mathrm{Res}(A, B)$. Let $n$ be the total number of variables in $D$, and $k$ be the width of $C$. For every $t \geq 1$, let $R_0, \ldots, R_{t-1}$ denote the $t$ consecutive complete rounds of the algorithm started with $D$, and let $D_0, \ldots, D_{t-1}$ denote the intermediate sets of clauses. Then, the probability that none of the $R_i$ is beneficial for $C$ and none of the $D_i$ absorbs $C$ is at most $e^{-t/(4n^k)}$.*

*Proof.* Let $R_0, \ldots, R_{t-1}$ denote the $t$ consecutive complete rounds of the algorithm started with $D$, and let $D_0, \ldots, D_{t-1}$ be the intermediate sets of clauses. In particular $D_0 = D$ and $R_i$ is a round started with $D_i$. For every $i \in \{0, \ldots, t-1\}$ let $\mathcal{R}_i$ be the event that $R_i$ is not beneficial and let $\mathcal{D}_i$ be the event that $D_i$ does not absorb $C$. We want to compute an upper bound for the joint probability of these events. Note that

$$\Pr\left[\bigcap_{i=0}^{t-1} \mathcal{R}_i \cap \mathcal{D}_i\right] = \prod_{j=0}^{t-1} \Pr\left[\mathcal{R}_j \cap \mathcal{D}_j \;\Big|\; \bigcap_{i=0}^{j-1} \mathcal{R}_i \cap \mathcal{D}_i\right] \leq \prod_{j=0}^{t-1} \Pr\left[\mathcal{R}_j \;\Big|\; \mathcal{D}_j \cap \bigcap_{i=0}^{j-1} \mathcal{R}_i \cap \mathcal{D}_i\right] \quad (1)$$

Hence, we shall give appropriate upper bounds for the factors on the right hand side of this inequality. To do this, let us first bound $\Pr\left[\bar{\mathcal{R}}_j \mid \mathcal{D}_j, \mathcal{R}_{j-1}, \mathcal{D}_{j-1}, \ldots, \mathcal{R}_0, \mathcal{D}_0\right]$ from below. Under the conditions $\mathcal{D}_j, \mathcal{R}_{j-1}, \mathcal{D}_{j-1}, \ldots, \mathcal{R}_0, \mathcal{D}_0$, Lemma 10 implies that there is an inconclusive round $R$ started with $D_j$ which is beneficial for $C$ at some $x^a \in C$. The probability that $R_j$ is beneficial for $C$ is bounded from below by the probability that $R_j$ is beneficial for $C$ at $x^a$. We will therefore bound the latter from below.

First let us compute a lower bound on the probability that the first $k-1$ decisions of the decision strategy are chosen to falsify $C \setminus \{x^a\}$ and the $k$-th choice is $x \overset{\mathrm{d}}{=} \bar{a}$. The probability that these choices are made is at least

$$\left[\left(\frac{k-1}{2n}\right)\left(\frac{k-2}{2(n-1)}\right) \cdots \left(\frac{1}{2(n-k+2)}\right)\right]\left(\frac{1}{2(n-k+1)}\right) \geq \frac{(k-1)!}{2^k n^k} \geq \frac{1}{4n^k}.$$

Note that a round started with $D_j$ that follows these choices may not even be able to do some of the decisions as the corresponding assignments may be implied. However, before the decision $x \overset{\mathrm{d}}{=} \bar{a}$ is made, a round following these choices will only perform decisions that agree with $R$ in $C \setminus \{x^a\}$ and therefore stay subsumed by $R$ after every new decision, by Lemma 1. In particular, right before the decision $x \overset{\mathrm{d}}{=} \bar{a}$ it will be inconclusive, it will falsify $C \setminus \{x^a\}$, and it will leave $x$ unset. Also by Lemma 1 it has performed the same assignments as $R$ up to order, and therefore the addition of $x \overset{\mathrm{d}}{=} \bar{a}$ will make it conclusive. It follows that the probability that the round will be beneficial for $C$ at $x^a$ can only be bigger.

Consequently, the probability of $\mathcal{R}_j$ conditional on $\mathcal{D}_j, \mathcal{R}_{j-1}, \mathcal{D}_{j-1}, \ldots, \mathcal{R}_0, \mathcal{D}_0$ is bounded from above by $1 - \frac{1}{4n^k}$. Therefore, by equation (1) we have

$$\Pr\left[\bigcap_{i=0}^{t-1} \mathcal{R}_i \cap \mathcal{D}_i\right] \leq \left(1 - \frac{1}{4n^k}\right)^t \leq e^{-t/(4n^k)}$$

where in the second inequality we used the fact that $1 + x \leq e^x$ for every real number $x$. $\quad\square$

## 3.5 Bounds

With the tools given above, we are now able to prove the main result of the paper: the simulation of width-$k$ resolution by the algorithm. We shall first give the proof for the algorithm employing the DECISION learning scheme. Not only is the proof easier and more instructive, but also we get slightly better bounds for this special case. Afterwards, we will see the result for asserting learning schemes in general.

### 3.5.1 THE DECISION SCHEME

The fact that makes DECISION easier to analyze is that, for this learning scheme, the occurrence of a beneficial round immediately yields absorption at the next step. Indeed, if $R$ is beneficial for $C$, then it branches in $C$, which means that the clause learned in this complete round is a subset of $C$. In particular this means that the next set of clauses will absorb a subset of $C$, and hence $C$ as well by Lemma 5. We obtain the following result as a direct consequence to Lemma 11.

**Lemma 12.** *Let $D$ be a set of clauses, and let $A$ and $B$ be two resolvable clauses that are absorbed by $D$ and that have a non-empty resolvent $C = \mathrm{Res}(A, B)$. Let $n$ be the total number of variables in $D$ and $k$ be the width of $C$. Then, for all $t \geq 1$, using the DECISION learning scheme, the probability that $C$ is not absorbed by the current set of clauses after $t$ restarts is at most $e^{-t/(4n^k)}$.*

*Proof.* Let $R_0, \ldots, R_{t-1}$ denote the $t$ consecutive complete rounds of the algorithm started with $D$, and let $D_0, \ldots, D_t$ be the intermediate sets of clauses. In particular $D_0 = D$ and $R_i$ is a round started with $D_i$. For every $i \in \{0, \ldots, t-1\}$ let $\mathcal{R}_i$ be the event that $R_i$ is not beneficial for $C$ and let $\mathcal{D}_i$ be the event that $D_i$ does not absorb $C$. If one of the $R_i$ is beneficial for $C$, then $D_{i+1}$ absorbs $C$. To see this, note that as $R$ branches in $C$, the clause $C_i$ learned from $R_i$ satisfies $C_i \subseteq C$. Hence $D_{i+1}$ absorbs both $C_i$ and $C$ by Lemma 5. Further, $D_t$ also absorbs $C$, if one of the $D_i$ absorbs it again by Lemma 5. Hence, the probability that $C$ is not absorbed by $D_t$ is bounded from above by $\Pr[\bigcap_{i=0}^{t-1} \mathcal{R}_i \cap \mathcal{D}_i]$. Lemma 11 implies that this is bounded by $e^{-t/(4n^k)}$. $\quad\square$

**Theorem 13.** *Let $F$ be a set of clauses on $n$ variables having a resolution refutation of width $k$ and length $m$. With probability at least $1/2$, the algorithm started with $F$, using the* Decision *learning scheme, learns the empty clause after at most $4m\ln(4m)n^k$ conflicts and restarts.*

*Proof.* The resolution refutation must terminate with an application of the resolution rule of the form $\text{Res}(x, \bar{x})$. We will show that for both $\ell = x$ and $\ell = \bar{x}$, the probability that $\{\ell\}$ is not absorbed by the current set of clauses after $4m\ln(4m)n^k$ restarts is at most $1/4$. Thus, both $\{x\}$ and $\{\bar{x}\}$ will be absorbed with probability at least $1/2$. If this is the case, it is straightforward that every complete round of the algorithm is conclusive. In particular, the round that does not make any decision is conclusive, and in such a case the empty clause is learned.

Let $C_1, C_2, \ldots, C_r = \{\ell\}$ be the resolution proof of $\{\ell\}$ that is included in the width-$k$ resolution refutation of $F$. In particular $r \leq m - 1$ and every $C_i$ is non-empty and has width at most $k$. Let $D_0, D_1, \ldots, D_s$ be the sequence of clause-sets produced by the algorithm where $s = rt$ and $t = \lceil 4\ln(4r)n^k \rceil$. For every $i \in \{0, \ldots, r\}$, let $\mathcal{E}_i$ be the event that every clause in the initial segment $C_1, \ldots, C_i$ is absorbed by $D_{it}$, and let $\overline{\mathcal{E}_i}$ be its negation. Note that $\Pr[\mathcal{E}_0] = 1$ vacuously and hence $\Pr[\overline{\mathcal{E}_0}] = 0$. For $i > 0$, we bound the probability that $\mathcal{E}_i$ does not hold conditional on $\mathcal{E}_{i-1}$ by cases. Let $p_i = \Pr[\overline{\mathcal{E}_i} \mid \mathcal{E}_{i-1}]$ be this probability. If $C_i$ is a clause in $F$, we have $p_i = 0$ by Lemma 5. If $C_i$ is derived from two previous clauses, we have $p_i \leq e^{-t/(4n^k)}$ by Lemma 12, which is at most $1/(4r)$ by the choice of $t$.

The law of total probability gives

$$\Pr\left[\overline{\mathcal{E}_i}\right] = \Pr\left[\overline{\mathcal{E}_i} \mid \mathcal{E}_{i-1}\right]\Pr[\mathcal{E}_{i-1}] + \Pr\left[\overline{\mathcal{E}_i} \mid \overline{\mathcal{E}_{i-1}}\right]\Pr\left[\overline{\mathcal{E}_{i-1}}\right]$$
$$\leq \Pr\left[\overline{\mathcal{E}_i} \mid \mathcal{E}_{i-1}\right] + \Pr\left[\overline{\mathcal{E}_{i-1}}\right].$$

Adding up over all $i \in \{1, \ldots, r\}$, together with $\Pr[\overline{\mathcal{E}_0}] = 0$, gives $\Pr[\overline{\mathcal{E}_r}] \leq \sum_{i=1}^{r} p_i \leq \frac{r}{4r} = \frac{1}{4}$. Since the probability that $C_r$ is not absorbed by $D_{rt}$ is bounded by $\Pr[\overline{\mathcal{E}_r}]$, the proof follows. $\square$

### 3.5.2 Asserting Learning Schemes in General

We shall now study the algorithm applying an arbitrary asserting learning scheme. The analysis is a bit more complex than that of the Decision scheme since in general a clause learned from a complete round $R$ cannot be assumed to be a subset of the decisions in $R$. Therefore we can only show that the resolvent is eventually absorbed by a little detour. We note that this proof has to overcome similar difficulties as, and is inspired by[3], the proof of Proposition 2 in the work of Pipatsrisawat and Darwiche (2009).

We need some preparation. Let $C$ be a clause and $D$ be a set of clauses. Let $W_{C,D}$ denote the set of literals $\ell$ in $C$ such that there exists an inconclusive round started with $D$ that is beneficial for $C$ at $\ell$. Let $u_{\ell,C,D}$ denote the number of variables left unassigned by an inconclusive round started with $D$ which is beneficial for $C$ at $\ell$. If no such round exists, we define $u_{\ell,C,D} = 0$. Note that this number is well-defined, as it follows easily from

---

3. We thank an anonymous reviewer for pointing out that the original proof of Proposition 2 in the work of Pipatsrisawat and Darwiche (2009) contained an error that was corrected in the version of the paper on their webpage. Our proof is not affected by this error.

Lemma 1 that every inconclusive round started with $D$ which is beneficial for $C$ at $\ell$ leaves the same number of variables unassigned. Further, define

$$u_{C,D} = \sum_{\ell \in W_{C,D}} u_{\ell,C,D}.$$

Note that if $C$ is absorbed by $D$, then $W_{C,D} = \emptyset$. Moreover, under the hypothesis of Lemma 10, the converse is also true. Analogously, if $C$ is absorbed by $D$, then $u_{C,D} = 0$ and, under the hypothesis of Lemma 10, the converse is also true.

**Lemma 14.** *Let $D$ and $D'$ be sets of clauses with $D \subseteq D'$. Let $A$ and $B$ be two resolvable clauses that are absorbed by $D$, and let $C = \mathrm{Res}(A, B)$. Then, $W_{C,D'} \subseteq W_{C,D}$ and $u_{\ell,C,D'} \leq u_{\ell,C,D}$ for all $\ell \in W_{C,D}$.*

*Proof.* If $W_{C,D'} = \emptyset$, nothing is to be shown. Otherwise, for $x^a$ in $W_{C,D'}$, we start by showing that $x^a$ belongs to $W_{C,D}$. Let $R'$ be an inconclusive round started with $D'$ which is beneficial for $C$ at $\ell$. Application of Lemma 2 to $R'$ and $C \backslash \{x^a\}$ yields an inconclusive round $R$ started with $D$ with the following properties: $R'$ subsumes $R$, both agree on $C \backslash \{x^a\}$, and $R$ branches in $C \backslash \{x^a\}$. To show that $R$ is beneficial for $C$ at $x^a$, it only remains to prove that extending $R$ by $x \stackrel{\mathrm{d}}{=} \bar{a}$ yields a conclusive round. Let $R^*$ be a round defined by this extension. Let $y$ be such that $C = \mathrm{Res}(A, B, y)$. Then $R^*$ falsifies $B \backslash \{\bar{y}\}$ and $A \backslash \{y\}$. By absorption, $R^*$ cannot be inconclusive, as otherwise, $y$ and $\bar{y}$ would be satisfied by $R^*$. This proves $W_{C,D'} \subseteq W_{C,D}$.

Now, we show that $u_{\ell,C,D'} \leq u_{\ell,C,D}$ for every $\ell$ in $W_{C,D}$. If $\ell$ does not belong to $W_{C,D'}$ nothing is to be shown since $u_{\ell,C,D'} = 0$ in that case. Otherwise, let $R'$ and $R$ be inconclusive rounds beneficial for $C$ at $\ell$ such that $R'$ is started with $D'$ and $R$ is started with $D$. By Lemma 1, $R'$ subsumes $R$, which finishes the proof. $\qquad \square$

**Lemma 15.** *Let $D$ be a set of clauses, let $A$ and $B$ be two resolvable clauses that are absorbed by $D$, and let $C = \mathrm{Res}(A, B)$. Let $R$ be a conclusive round started with $D$ and let $D'$ be obtained from $D$ by adding the asserting clause learned from $R$. If $C$ is not empty and $R$ is beneficial for $C$ at some $\ell \in C$, then $u_{\ell,C,D'} < u_{\ell,C,D}$ and $u_{C,D'} < u_{C,D}$.*

*Proof.* By Lemma 14 we already know that $u_{C,D'} \leq u_{C,D}$ and $u_{\ell,C,D'} \leq u_{\ell,C,D}$. Therefore, it suffices to demonstrate that, in the presence of $R$, the second inequality is strict.

By hypothesis, $R$ is beneficial for $C$ at $\ell$. Let $C'$ be the asserting clause learned by $R$. Let $R^*$ be the unique inconclusive round contained in $R$ which is beneficial for $C$ at $\ell$; this is the round which does not contain the last decision made by $R$. By Lemma 1, the number of assignments made by any two rounds started with $D$ and beneficial for $C$ at $\ell$ are the same. Hence, the number of variables left unassigned by $R^*$ equals $u_{\ell,C,D}$, and $u_{\ell,C,D} \geq 1$ since at least one variable is unset.

If $u_{\ell,C,D'} = 0$ then already $u_{\ell,C,D'} < u_{\ell,C,D}$. Therefore, assume that $u_{\ell,C,D'} \geq 1$. In particular, there exists an inconclusive round $R'$ started with $D'$ which is beneficial for $C$ at $\ell$. By Lemma 1 the round $R'$ subsumes $R^*$. By the definition of asserting clauses, $C'|_{R^*}$ is a unit clause, and since $C'$ belongs to $D'$, it is absorbed by $D'$ and hence $R'$ satisfies $C'$. This proves that $R'$ sets at least one more variable than $R^*$ and therefore $u_{\ell,C,D'} < u_{\ell,C,D}$. $\qquad \square$

With these two technical lemmas in hand we are ready to state and prove the analogue of Lemma 12 for arbitrary asserting learning schemes.

**Lemma 16.** *Let $D$ be a set of clauses, and let $A$ and $B$ be two resolvable clauses that are absorbed by $D$ and that have a non-empty resolvent $C = \mathrm{Res}(A, B)$. Let $n$ be the total number of variables in $D$ and let $k$ be the width of $C$. Then, for all $t \geq 1$, using an arbitrary asserting learning scheme, the probability that $C$ is not absorbed by the current set of clauses after $kn \cdot t$ restarts is at most $kn \cdot e^{-t/(4n^k)}$.*

*Proof.* Let $b = u_{C,D}$, and $s = bt$, and let $D_0, \ldots, D_s$ be the sequence of sets of clauses produced by the algorithm, starting with $D_0 = D$. For every $i \in \{0, \ldots, b\}$, let $X_i = u_{C,D_{it}}$ and let $\mathcal{E}_i$ be the event that $X_i \leq b - i$.

We will bound the probability that $C$ is not absorbed by $D_{bt}$ from above. Since this event implies that $X_b \neq 0$, it suffices to bound $\Pr[\,\overline{\mathcal{E}_b}\,]$. Note that $\Pr[\,\mathcal{E}_0\,] = 1$ vacuously and hence $\Pr[\,\overline{\mathcal{E}_0}\,] = 0$. For $i > 0$, we bound the probability that $\mathcal{E}_i$ does not hold. The law of total probability gives

$$\Pr\left[\,\overline{\mathcal{E}_i}\,\right] = \Pr\left[\,\overline{\mathcal{E}_i} \mid \mathcal{E}_{i-1}\,\right] \Pr\left[\,\mathcal{E}_{i-1}\,\right] + \Pr\left[\,\overline{\mathcal{E}_i} \mid \overline{\mathcal{E}_{i-1}}\,\right] \Pr\left[\,\overline{\mathcal{E}_{i-1}}\,\right]$$
$$\leq \Pr\left[\,\overline{\mathcal{E}_i} \mid \mathcal{E}_{i-1}\,\right] + \Pr\left[\,\overline{\mathcal{E}_{i-1}}\,\right].$$

Let $p_i = \Pr[\,\overline{\mathcal{E}_i} \mid \mathcal{E}_{i-1}\,]$ and note that $\Pr[\,\overline{\mathcal{E}_i} \mid X_{i-1} < b - i + 1\,] = 0$. Hence we have $p_i \leq \Pr[\,\overline{\mathcal{E}_i} \mid X_{i-1} = b - i + 1\,]$. Consider the sequence $D_{(i-1)t+1}, \ldots, D_{it}$ of sets of clauses and the corresponding complete rounds of the algorithm. Conditional on $X_{i-1} = b - i + 1$, the event $\overline{\mathcal{E}_i}$ implies that $X_i = X_{i-1} \neq 0$ and hence none of the above sets of clauses absorbs $C$. Furthermore, by Lemma 15, none of the corresponding rounds is beneficial for $C$. Thus, by Lemma 11, we have $p_i \leq e^{-t/(4n^k)}$. Adding up over all $i \in \{1, \ldots, r\}$, together with $\Pr[\,\overline{\mathcal{E}_0}\,] = 0$, gives $\Pr\left[\,\overline{\mathcal{E}_b}\,\right] \leq \sum_{i=1}^{b} p_i \leq b \cdot e^{-t/(4n^k)}$. The Lemma follows as necessarily $b \leq kn$. □

We are now able to prove the main theorem.

**Theorem 17.** *Let $F$ be a set of clauses on $n$ variables having a resolution refutation of width $k$ and length $m$. With probability at least $1/2$, the algorithm started with $F$, using an arbitrary asserting learning scheme, learns the empty clause after at most $4km \ln(4knm)n^{k+1}$ conflicts and restarts.*

*Proof.* The proof is analogous to the proof of Theorem 13 with Lemma 16 playing the role of Lemma 12, and choosing $t = \lceil 4\ln(4m \cdot kn)n^k \rceil$ now. □

## 4. Consequences

The total number of clauses of width $k$ on $n$ variables is bounded by $2^k \binom{n}{k}$, which is at most $2n^k$ for every $n$ and $k$. Therefore, if $F$ has $n$ variables and a resolution refutation of width $k$, we may assume that its length is at most $4n^k$ by the following estimate

$$\sum_{i=0}^{k} 2^i \binom{n}{i} \leq 1 + 2\sum_{i=1}^{k} n^i = 1 + 2n \cdot \left(\frac{n^k - 1}{n - 1}\right) \leq 4n^k.$$

We obtain the following consequence to Theorem 17.

**Corollary 18.** *Let $F$ be a set of clauses on $n$ variables having a resolution refutation of width $k$. With probability at least $1/2$, the algorithm started with $F$, using an arbitrary asserting learning scheme, learns the empty clause after at most $16k(k+1)\ln(16kn)n^{2k+1}$ conflicts and restarts.*

An application of Corollary 18 is that, even though it is not explicitly defined for the purpose, the algorithm can be used to decide the satisfiability of CNF formulas of treewidth at most $k$ in time $O(k^2 \log(kn)n^{2k+3})$. This follows from the known fact that every unsatisfiable formula of treewidth at most $k$ has a resolution refutation of width at most $k+1$ (Alekhnovich & Razborov, 2002; Dalmau, Kolaitis, & Vardi, 2002; Atserias & Dalmau, 2008).

If we are interested in producing a satisfying assignment when it exists, we proceed by self-reducibility: we assign variables one at a time, running the algorithm $\log_2(n)+1$ times after each assignment to detect if the current partial assignment cannot be extended any further, in which case we choose the complementary value for the variable. For this we use the fact that if $F$ has treewidth at most $k$, then $F|_{x=a}$ also has treewidth at most $k$. For the analysis, note that since each run of the algorithm is correct with probability at least $1/2$, each new assignment is correct with probability at least

$$1 - 2^{-(\log_2(n)+1)} = 1 - \frac{1}{2n}.$$

This means that all iterations are correct with probability at least $(1 - \frac{1}{2n})^n \geq \frac{1}{2}$. The running time of this algorithm is $O(k^2(\log(kn))^2 n^{2k+4})$.

## Acknowledgments

We thank Martin Grohe for suggesting the problem of comparing the power of SAT-solvers with bounded-width resolution. We also thank Knot Pipatsrisawat and Adnan Darwiche for pointing out the connection between 1-empowering and absorption. Thanks also to Peter Jeavons for comments on the conference version of this paper, and to the anonymous referees for very detailed comments.

The first author was supported in part by CYCIT TIN2007-68005-C04-03. The second author was supported in part by the European Research Council (ERC), Grant 239962. The third author was supported in part by a fellowship within the Postdoc-Programme of the German Academic Exchange Service (DAAD). A preliminary version of this paper appeared in the Proceedings of the 12th International Conference on Theory and Applications of Satisfiability Testing, SAT'09 (Atserias et al., 2009).